\documentclass[useAMS,usenatbib]{mn2e}

\usepackage{url,times,graphicx,amsmath,amsfonts,amssymb,aas_macros,color,epsfig,epstopdf}

\newcommand{\mockalph}[1]{}
\newcommand{\hMpc}{{\ifmmode{h^{-1}{\rm Mpc}}\else{$h^{-1}$Mpc}\fi}}
\newcommand{\hkpc}{{\ifmmode{h^{-1}{\rm kpc}}\else{$h^{-1}$kpc}\fi}}
\newcommand{\hMsun}{{\ifmmode{h^{-1}{\rm {M_{\odot}}}}\else{$h^{-1}{\rm{M_{\odot}}}$}\fi}}

\newcommand{\gtsima}{$\; \buildrel > \over \sim \;$}
\newcommand{\gsim}{\lower.5ex\hbox{\gtsima}}

\def\lesssim{\mathrel{\hbox{\rlap{\hbox{\lower4pt\hbox{$\sim$}}}\hbox{$<$}}}}
\def\gtrsim{\mathrel{\hbox{\rlap{\hbox{\lower4pt\hbox{$\sim$}}}\hbox{$>$}}}}
\def\lsim{\mathrel{\rlap{\lower3pt\hbox{$\sim$}}
    \raise1pt\hbox{$<$}}}                
\def\gsim{\mathrel{\rlap{\lower3pt\hbox{$\sim$}}
    \raise1pt\hbox{$>$}}}                

\newcommand{\mstar}{M$_\star$}
\newcommand{\lcdm}{$\Lambda$CDM}

\newcommand{\vmax}{V$_{\rm max}$}
\newcommand{\vrone}{V$_{\rm rot}$(1kpc)}
\newcommand{\vrot}{V$_{\rm rot}$}

\newcommand{\msun}{M$_\odot$}

\newcommand{\mhalo}{M$_{\rm halo}$}

\newcommand{\beq}{\begin{equation}}
\newcommand{\eeq}{\end{equation}}
\def\beqa{\begin{eqnarray}}
\def\eeqa{\end{eqnarray}}
\def\hMpc{$h^{-1}\,{\rm Mpc}$}
\def\hkpc{$h^{-1}\,{\rm kpc}$}
\def\kpc{\,kpc}

\def\kms{km$\,$s$^{-1}$}

\def\head{
 \vbox to 0pt{\vs.
                   \hbox to 0pt{\hskip 440pt\rm LA-UR-10-07069\hss}
                  \vskip 25pt}}

\title[Rotation Curve Variation]{The Variation of Rotation Curve Shapes as a Signature of the Effects of Baryons on Dark Matter Density Profiles}
\author[Brook ]{Chris B. Brook$^{1,2}$
\thanks{E-mail:cbabrook@gmail.com}\\
$^{1}$Departamento de F\'isica Te\'orica, M\'odulo C-15, Facultad de Ciencias, Universidad Aut\'onoma de Madrid, 28049 Cantoblanco, Madrid, Spain\\
$^{2}$Ramon y Cajal Fellow\\
}

\begin{document}

\date{Accepted XXXX . Received XXXX; in original form XXXX}

\pagerange{\pageref{firstpage}--\pageref{lastpage}} \pubyear{2010}

\maketitle

\label{firstpage}


\begin{abstract}
Rotation curves of  galaxies show a wide range of shapes, which can be paramaterized as scatter in \vrone/\vmax\, i.e. the ratio of the rotation velocity measured at 1$\,$kpc and the maximum measured rotation velocity. We examine whether the observed scatter can be accounted for by  combining scatters in disc scale-lengths, the concentration-halo mass relation, and the \mstar-\mhalo\ relation. We use  these scatters to create model galaxy populations; when  housed within dark matter halos that have  universal, NFW density profiles, the model does not match the lowest observed values of \vrone/\vmax\,  and has too  little scatter in \vrone/\vmax\ compared to observations. By contrast, a model using a mass dependent dark matter profile, where the inner slope is determined by the ratio of \mstar/\mhalo, produces galaxies with low values of  \vrone/\vmax\, and a much larger scatter, both in  agreement with observation. We conclude that the large  observed scatter in \vrone/\vmax\ favours density profiles that are significantly affected by baryonic processes.  
 Alternative dark matter core formation models such as SIDM may also account for the observed variation in rotation curve shapes, but these observations may provide important constraints in terms of  core sizes, and whether they vary with halo mass and/or merger history.
\end{abstract}
\noindent
\begin{keywords}
 methods: galaxies: formation - haloes
 \end{keywords}
   
\section{Introduction} \label{sec:introduction}

The rotation curves of disc galaxies are diverse \citep[e.g.][]{zwaan95,swaters09}; even galaxies with similar velocities at the flat part of the rotation curve can have significantly different rotation velocities at small radii. This results in a large scatter in rotation velocity at small radii, when plotted as a function of \vmax\ (\citealt{oman15} who measure \vrot\ at 2\,kpc). 

It is well documented that a population of galaxies have slowly rising rotation curves, meaning a low value of \vrone\ compared to \vmax. Such galaxies are presented as evidence for dark matter profiles that are relatively flat in the inner region,  often referred to as cored profiles \citep{moore94,salucci03,denaray08,oh11}. It is argued that this contradicts a prediction of cold dark matter cosmology, that dark mater halos have steep inner density profiles, or ``cusps" \citep{navarro96}. Yet other galaxies have rotation curves that do rise steeply in the inner region, and which can be fit by cuspy dark matter density profiles, having relatively high values of \vrone\ compared to \vmax. It would seem that  a successful model of galaxy formation needs to account for both cored and cuspy density profiles, sometimes in galaxies that have  similar dynamical masses. 

This Letter examines the scatter in the  \vrone-\vmax\ plane in a  model population of  galaxies that incorporates the empirical scatter in disc scale-lengths \citep{courteau07,swaters09,fathi10}, scatter in the concentration-halo mass relation \citep{dutton14}, and scatter in the \mstar-\mhalo\ relation \citep{reddick13,behroozi13}. Different forms of the underlying dark matter halo are assumed. Firstly the universal ``NFW"  density profile that provides a good fit to dark matter only simulations \citep{navarro96}. Secondly a mass dependent density profile that accounts for the effects of baryons on dark matter (\citealt{dicintio14a}a,b). And finally, a cored density profile that may occur in alternative dark matter theories, in particular self-interacting dark matter (SIDM). 

\begin{figure}
\hspace{-.cm}
  \includegraphics[width=3.in]{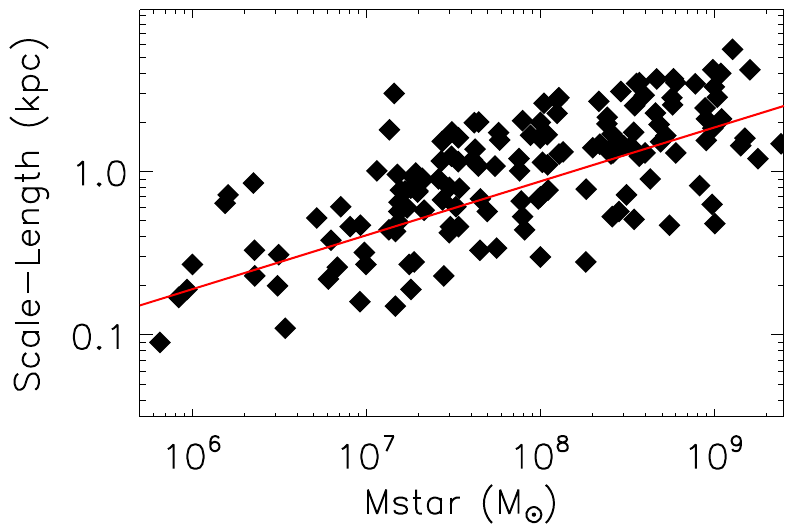}
  \caption{Scale length  versus luminosity for  144 observed galaxies (Hunter \& Elmegreen 2006),  which is fit by the red line,  around which there is a scatter of  $\sim$0.23$\,$dex}
\label{fig:scatterh} 
\end{figure}

In Section~\ref{model}, we present our  model galaxy populations. In Section~\ref{observations} we compile  an observational sample of galaxies which have detailed rotation curve data. In Section~\ref{results} we show the model populations in the \vrone-\vmax\ plane, and plot the scatter in    \vrone/\vmax\ as a function of \vmax, making comparisons with the observational sample. We discuss our findings in Section~\ref{discussion}, concluding that baryonic physics  play a role in shaping the dark matter density profiles in the inner 1$\,$kpc of galaxies.


  \section{Method}\label{method}

\subsection{Model}\label{model}
Our model galaxies consist of exponential stellar and gas discs embedded within dark matter halos. 
For the dark matter halos, we use  the concentration-mass relation for Planck \lcdm\ \citep{dutton14}, including a correction for adiabatic contraction, as found in cosmological simulations \citep{dicintio14b}. This  contraction correction  alters the concentration rather than the inner slope, allowing the inner slope of the density profile  to be set by hand, and has the form:
\begin{equation}
{\rm C}=(1.0+ 0.00003e^{3.4X})\times {\rm C}_{\rm NFW}\\
 \label{c}
\end{equation}
where $X=\log_{10}$(\mstar/\mhalo) + 4.5, and C$_{\rm NFW}$ is the concentration given from the adopted  \citep{dutton14} relation.\\
\vspace{-.35cm}

Galaxy stellar masses are matched to dark matter halos   using the empirical abundance matching relation \citep{guo10}. We  include the universal dark matter fraction of mass ($\Omega_{\rm DM}$/$\Omega_{m}$=0.83) in dark matter halos, and add a stellar disc with scale-lengths (h$_{s}$) taken from the observed \mstar-h$_s$ relation, log$_{10}$(h$_{\rm s}$)=$-$2.8+0.35log$_{10}$(\mstar), (Dutton 07 gives log$_{10}$(h$_{\rm s}$)=$-$2.46+0.281log$_{10}$(\mstar)) coming from a fit to galaxies from \cite{hunter06}, which is similar to the relation found previously in larger studies of generally more massive galaxies \citep{courteau07,fathi10}. We also add a gas  disk  with  mass  M$_{\rm gas}$$=$1.3M$_{\rm HI}$, with  log$_{\rm 10}$(M$_{\rm HI}$/\mstar)$=$$-0.43$log$_{10}($\mstar$)+3.75$ \citep{papastergis12} 
 and  disc scale length h$_{g}$=3h$_{s}$.

The mass dependent `DC14' density profile   is  based on cosmological simulations which match a wide range of galaxy scaling relations \citep{brook12,stinson13,kannan14,obreja14}. The DC14 profile  accounts for the expansion of dark matter halos due to  the effects of feedback from star forming regions \citep{dicintio14b}, taking the form \citep{merritt06}:
\begin{equation}
\rho(r)=\frac{\rho_s}{\left(\frac{r}{r_s}\right)^{\gamma}\left[1 + \left(\frac{r}{r_s}\right)^{\alpha}\right] ^{(\beta-\gamma)/\alpha}}
\label{five}
\end{equation}
\noindent where ($\alpha,\beta,\gamma$) are the sharpness of the transition, the outer and the inner slope, respectively. 

The DC14 profile allows for a  range of inner slopes,  determined by the stellar-to-halo mass ratio.   Galaxies with \mstar$\lsim$3$\times$10$^6$M$_\odot$ are dark matter dominated, and do not produce enough energy to flatten the halo's inner density profile, which remains steep \citep{penarrubia12,governato12,dicintio14a}.  The inner density profile then becomes increasingly flat as stellar mass increases relative to the dark matter mass \citep{governato12,dicintio14a}, with the greatest flattening  when   \mstar$\sim$3$\times$10$^8$\msun.  For higher masses, the deeper potential well is able to  oppose the halo expansion (\citealt{dicintio14a}a,b), resulting in a profile which becomes steeper, and returns to the NFW value at about the Milky Way mass.

 We make a simple parameterisation as representative of the   self interacting dark matter (SIDM) case, which is loosely based on \cite{zavala13}.  Core size is set to zero for Milky Way mass galaxies  (\mstar=6$\times$10$^{10}$\msun) and increases as the scale-length decreases, until reaching a maximum for \mstar$=$10$^8$\msun, with lower mass  halos all having the maximum core size, for which we use values of 1, 2 and 3$\,$kpc.  
 Cored profiles for the SIDM model have the form: 
\begin{equation}
 \rho(r)=\frac{\rho_0 r_s^3}{(r_c+r)(r_s+r)^2}
 \end{equation}
 \vspace{-.35cm}
 
\noindent where $\rho_0$ is a characteristic halo density, $r_s$ is a scale radius and $r_c$ is the core size \citep{penarrubia12}.  
 
 Note that this definition of core size,  set by the parameter $r_c$,  does not necessarily match definitions used in studies which use different density profiles, so care needs to be taken when comparing core sizes quoted here to those quoted in  other studies.  
 
 The important distinction between the SIDM model and the DC14 model is that in the SIDM case, core size depends on halo mass, while in the DC14 model the core size depends on the ratio of \mstar/\mhalo. We have not included scatter in the relation between \mhalo\ and core-size which may occur in SIDM models, dependent on different merger histories. Such a scatter is apparent in high mass SIDM halos \cite{rocha13}, but has not been quantified for low mass SIDM halos.

 \begin{figure}
\hspace{-.6cm}
  \includegraphics[width=3.8in]{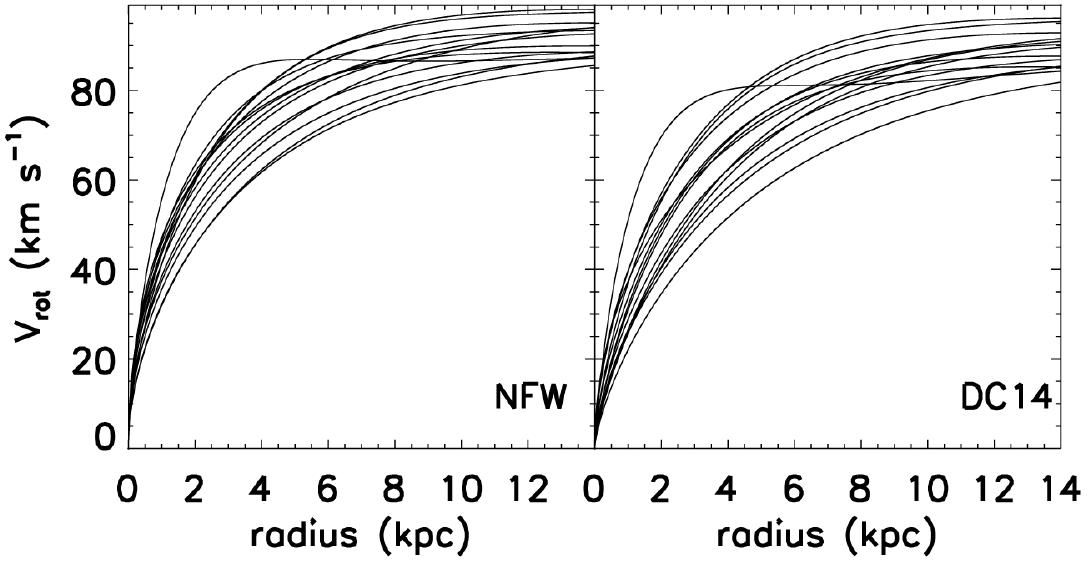}
  \caption{18 rotation curves are shown for the models with NFW and   DC14 density profiles in the left and right panels respectively. The total dark matter halo masses  associated with  this subset of model galaxies  have maximum velocities 90$<$V$_{\rm rot}^{\rm DM}$$<$100$\,$km$\,$s$^{-1}$.   }
\label{fig:rc} 
\end{figure}

\subsection{Scatter}
We then add scatter to our models. Scatter in disc scale length comes from the empirical scatter of 0.23$\,$dex in the luminosty-disc scale length relation,  determined using 144 observed galaxies  (\citealt{hunter06}, Oh et al 2015 in prep), as shown in Figure~\ref{fig:scatterh}. This is lower than the scatter of $\sim$0.3$\,$dex found in \cite{courteau07} for a larger sample of late type galaxies, but a sample which is of generally higher mass galaxies than used in this study. \cite{fathi10} explored the relation between scale-length and stellar mass down to 10$^{6.6}$\msun, and found similar scatter to  \cite{courteau07}, but  with  lower scatter for low mass galaxies.
We tested our results using a scatter of 0.3$\,$dex and found relatively insignificant effects that do not alter our conclusions. Scatter in the mass-concentration relation is set at  0.11$\,$dex \citep{dutton14}. Finally, scatter in \mstar-\mhalo\ is set at 0.2$\,$dex \citep{reddick13,behroozi13}.  

\begin{figure}
\hspace{-.1cm}
  \includegraphics[width=3.25in]{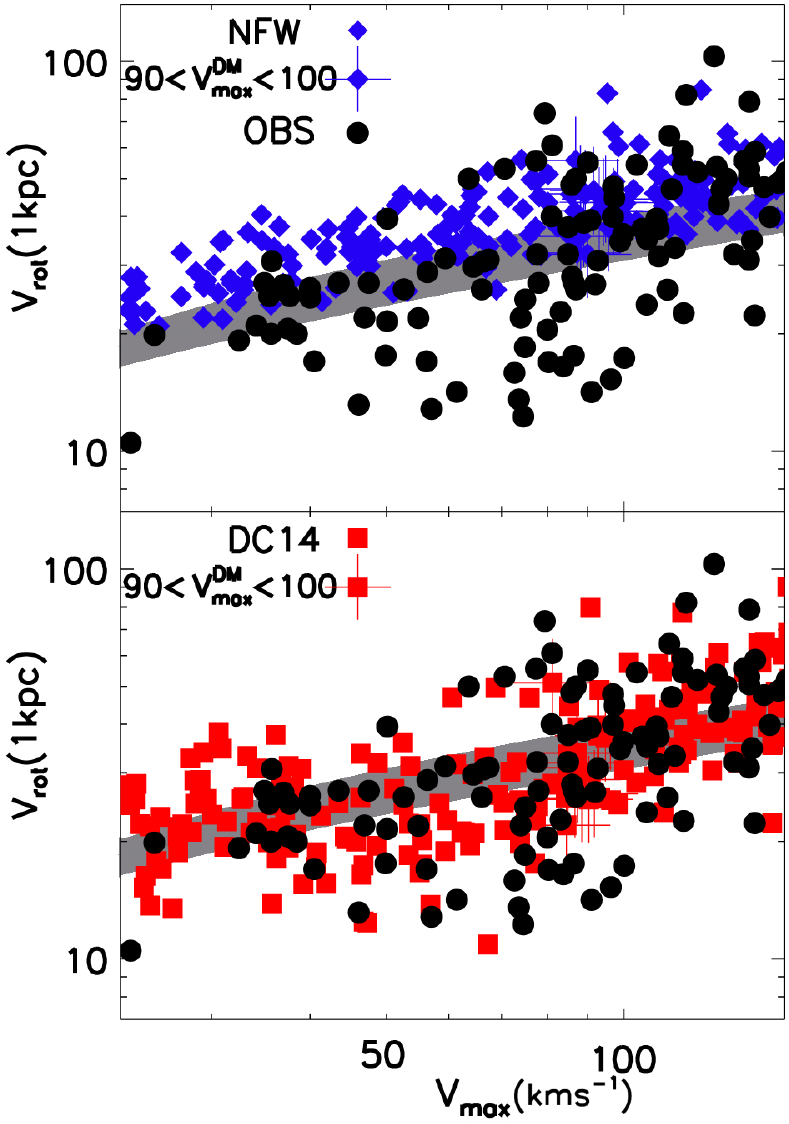}
  \caption{\vrone\ versus \vmax\ for the model galaxy  populations when assuming the NFW (top panel, blue diamonds) and DC14  (bottom panel, red squares) profiles. Black circles are observed galaxies. Values for halos from dark matter only   simulations, including 1$\sigma$ scatter, are shown as a grey band. The 18 galaxies shown in Figure~\ref{fig:rc} are marked with  '+' symbols. }
\label{fig:v1vmax} 
\end{figure}

\subsection{Observations}\label{observations}
We compiled  110 observed galaxies  (\citealt{deblok02,swaters03,deblok04,denaray08,deblok08,swaters09,oh11,mannheim12}; Oh et al 2015, and references therein) with  \vmax $<$150kms$^{-1}$ that have rotation curves that allow us to compare \vrone\ with  \vmax.

\section{Results}\label{results}

In Figure~\ref{fig:rc}, we show 18 rotation curves  for the models with NFW and   DC14 density profiles in the left and right panels respectively.  This subset of model galaxies have associated dark matter halos  with maximum velocities 90$<$V$_{\rm rot}^{\rm DM}$$<$100$\,$km$\,$s$^{-1}$: the only difference between the two panels are the density profiles. A significant range of rotation curve shapes is apparent in both models. The DC14  models have rotation curves that are more slowly rising than the NFW models.

\begin{figure}
\hspace{-.1cm}
  \includegraphics[width=3.25in]{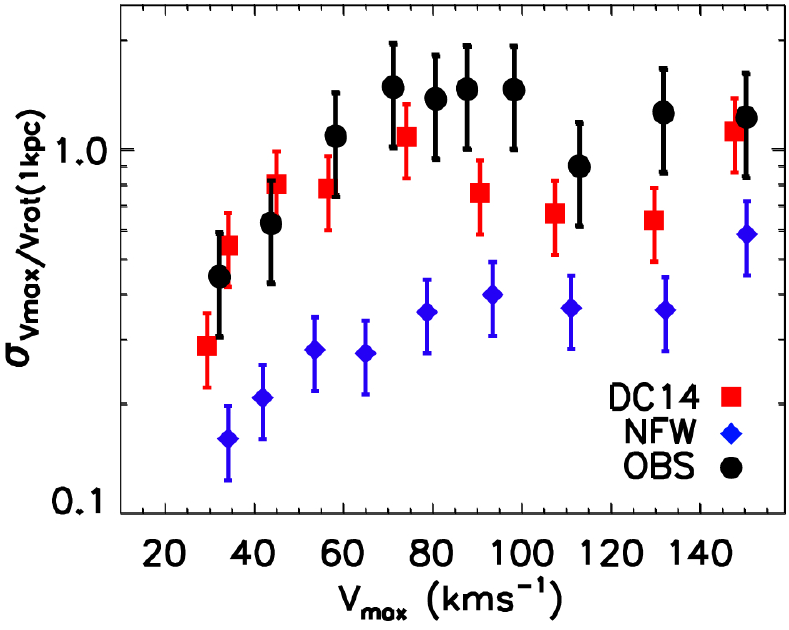}
  \caption{  Standard deviation of  the ratio \vrone/\vmax\ versus \vmax\ for the NFW  (blue diamonds) and DC14 (red squares) models. Observed galaxies are shown as black circles.  Error bars show the standard errors.}
\label{fig:sig} 
\end{figure}

To measure the diversity in rotation curve shapes we plot, in Figure~\ref{fig:v1vmax},  \vrone\  as a function  \vmax\ for the model galaxy  populations when assuming the NFW (top panel, blue diamonds) and DC14  (bottom panel, red squares) profiles. \vmax\  is defined as the rotation velocity as measured at 5 disc scale lengths, better representative of the observational data than simply plotting maximum circular velocity.  Also shown in each panel, as black circles,  are the observed galaxies. The value of \vmax\ for halos from dark matter only   simulations, including 1$\sigma$ scatter, are shown in each panel as a grey band. The 18 model galaxies whose rotation curves are shown in Figure~\ref{fig:rc} are marked with  '+' symbols.

Two features of Figure~\ref{fig:v1vmax} are worthy of attention. Firstly,  around  half the observed galaxies have values of \vrone/\vmax that fall below the NFW model population. This is just re-stating  that a significant number of  observed galaxies are better fit by cored profiles than cuspy profiles. By contrast, the DC14 model population includes galaxies  with low \vrone/\vmax, better matching the observations, 

The second feature is that the models with universal NFW profiles scatter around (and slightly above) the expected values from pure N-body simulations, with  no apparent shape to the distribution, in terms of mass dependance. This  reflects the universality of the NFW profile. By contrast, both the observed galaxies and the DC14 model galaxies appear to have  a structure in their distribution; a relatively flat relation at low velocities, with a positive relation between \vrone\ and \vmax\  for higher rotation velocities, i.e those with \vmax$\gsim$70$\,$\kms. 

We further explore the scatter in \vrone/\vmax\ in Figure~\ref{fig:sig}, where the standard deviation of  the ratio \vrone/\vmax\ is plotted as a function of \vmax\ for the NFW  (blue diamonds) and DC14 (red squares) models. The observed galaxies  are again shown as black circles.  In each case, error bars indicate the standard error. Clearly the NFW model has far too little scatter in the ratio  \vrone/\vmax\  at all values of \vmax. The DC14 model is a  better match to observations. The match is certainly not perfect, with the DC14 model having a peak dispersion at a slightly lower value of \vmax\ than is found in the observed population. 

\begin{figure}
\hspace{-.1cm}
  \includegraphics[width=3.25in]{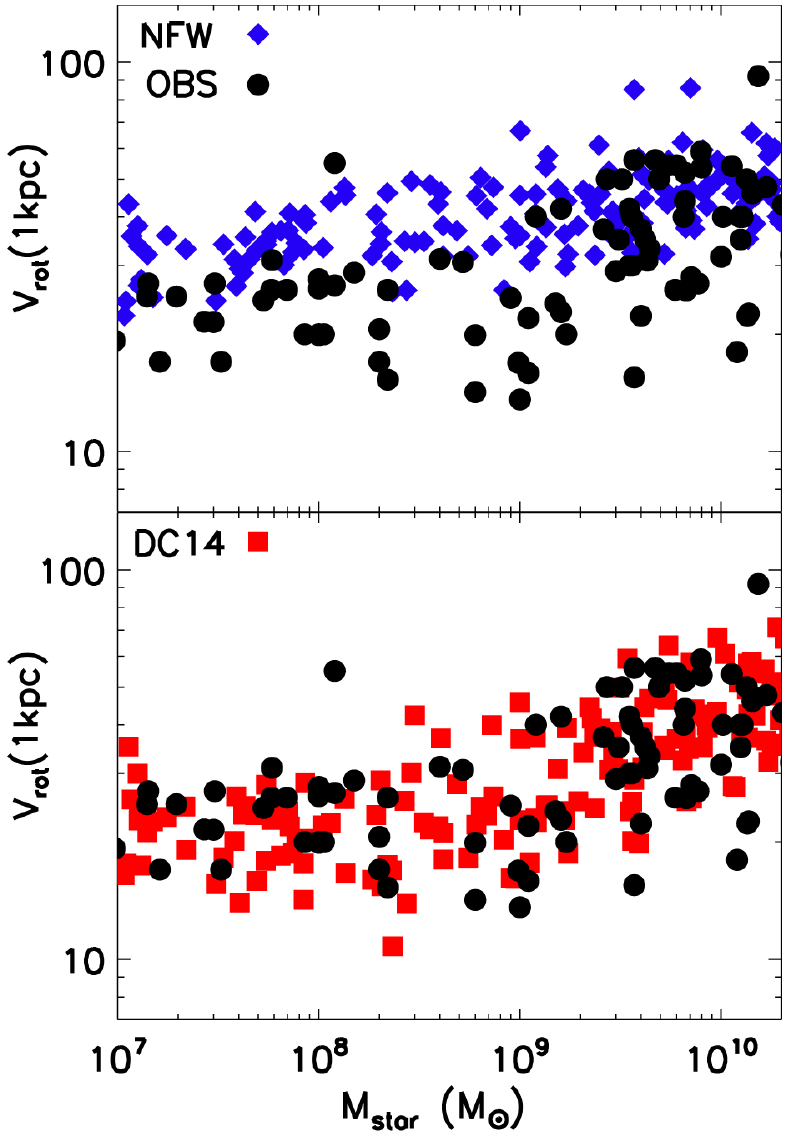}
  \caption{The relation between \mstar\ and V(1kpc), plotted for the NFW (blue diamonds, top panel) and DC14 (red squares, bottom panel) models, with observations shown as black circles.}
\label{fig:msv1} 
\end{figure}

\subsection{\mstar-V(1kpc)}\label{mv1}
 We now look at the relation between \mstar\ and V(1kpc), plotted for the NFW (top panel) and DC14 (bottom panel) models in Fig.~{fig:msv1}.  It is clear from Fig.~{fig:msv1} that the NFtW model is not able to form galaxies with low values of V(1kpc) at all stellar masses. By contrast, the observed relation between stellar mass and V(1kpc) is well produced by the DC14 model. We caution that the data set is not homogeneous in terms of  stellar masses. A homogeneous set of galaxy stellar masses, along with high resolution rotation curves, as being compiled by Lelli et al. 2015 (in prep), would be invaluable here.

\subsection{Self Interacting Dark Matter}\label{sidm}
\begin{figure}
\hspace{-.cm}
 \includegraphics[width=3.25in]{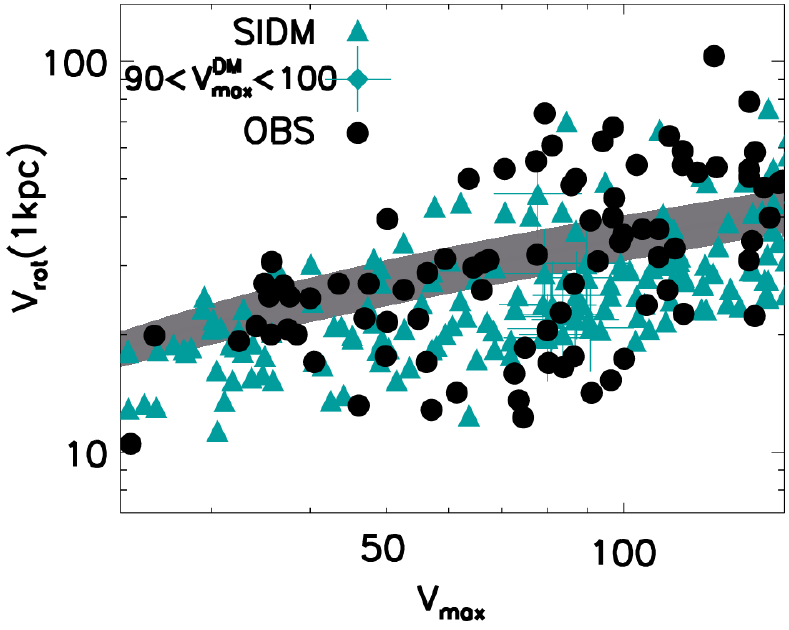}
  \caption{ \vrone\ versus \vmax\  for the SIDM model population with core size 3\kpc (green triangles) and observed galaxies (black circles). The expected values for halos from dark matter only   simulations, including 1$\sigma$ scatter, are shown as a grey band. Galaxies associated  with dark matter halos  with maximum velocities 90$<$V$_{\rm rot}^{\rm DM}$$<$100$\,$km$\,$s$^{-1}$  are marked with  '+' symbols}
\label{fig:v1vmaxS} 
\end{figure}

\begin{figure}
\hspace{-.1cm}
  \includegraphics[width=3.25in]{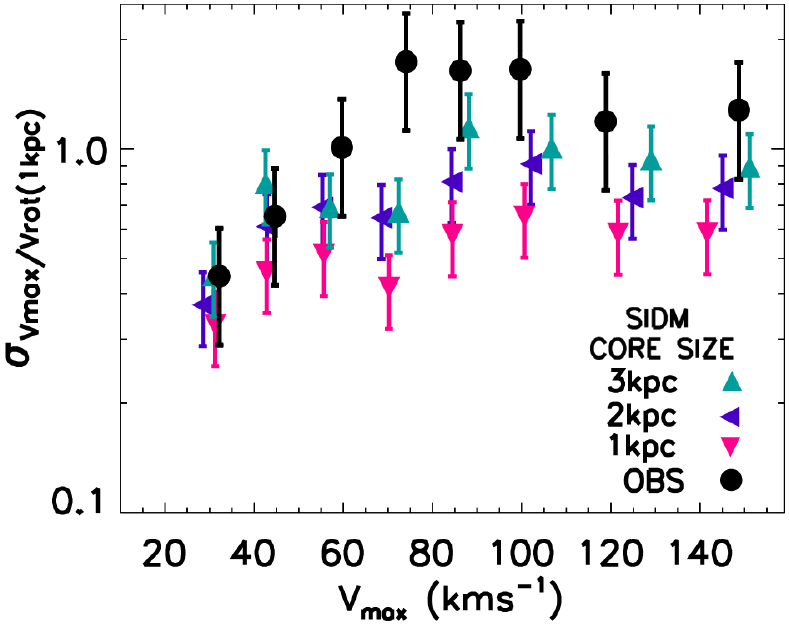}
  \caption{The scatter in the ratio \vrone/\vmax\, plotted as a function of \vmax\ for the SIDM model, shown for 1, 2 and 3 \kpc cores as pink,  violet, and green triangles respectively, as well as for the observed galaxies (black circles).  Error bars show the standard error: bins contain 11 (18) observed (model)  galaxies.}
\label{fig:sigS} 
\end{figure}

We made similar plots for the SIDM model population. Figure~\ref{fig:v1vmaxS} shows \vrone\ as a function of  \vmax\  as green  triangles, where we have used a maximum core size of 3$\,$kpc,  with  observed galaxies over-plotted as black circles. Again, expected values for dark matter only  simulations, including 1$\sigma$ scatter, are shown as a grey band. The galaxies associated  with dark matter halos  with maximum velocities 90$<$V$_{\rm rot}^{\rm DM}$$<$100$\,$km$\,$s$^{-1}$  are marked with  '+' symbols. Figure~\ref{fig:sigS} shows  the scatter in the ratio \vrone/\vmax\ plotted as a function of \vmax\ for the SIDM model for 3 different core sizes, 1\kpc (pink triangles), 2\kpc (purple triangles) and  3\kpc (green triangles) compared to  the observed galaxies.  

For the paramaterization that we have used, the SIDM model requires a relatively large core size, 3\kpc, in order to  reproduce the observed relations. As is evident in  Figure~\ref{fig:v1vmaxS},  SIDM model galaxies can have low values of \vrone, at given \vmax, as cored profiles make a better match for galaxies with slowly rising rotation curves. The SIDM models also have a larger range of rotation curve shapes than the NFW models.  The larger the cores in  SIDM halos, the greater is the effect of the scatter in the  baryonic mass distribution (i.e. scatter in scale-lengths)  in determining the variation in rotation curve shapes. The SIDM model with 3\kpc cores is able to do as well as the DC14 model in reproducing the observed scatter in the rotation curve shapes. 
 
What has not been included in the SIDM model is a dependence of core size on mass accretion history, which would result in  scatter in the core size at given halo mass. This is expected, and has been shown on cluster scales \cite{rocha13}, but has not been quantified at low masses. This scatter will likely make the SIDM model closer to observations, and difficult to distinguish from core formation through baryonic processes, although the detailed relation between scatter and galaxy mass may still differ between the models.

\section{Conclusions}\label{discussion}

We have created a model population of disc galaxies, using empirical relations between disc scale length and stellar mass, between stellar and gas mass, and between stellar and halo mass. Scatter in the \mstar-scale length and \mstar-\mhalo\ relations are added, also from empirical values. The model galaxies are housed within dark matter halos that follow the mass-concentration relation with scatter, as found in N-body simulations. We  compare model populations that make different assumptions for dark matter halo profiles, and explore the consequent relations between \vrone\ and \vmax.

When the model galaxy population is housed within halos with universal, steep, NFW density profiles,  none of the galaxies  have low  values of   V$_{\rm rot}$(1$\,$kpc)/\vmax,  as found in  some observed galaxies, and the model population has too little scatter in V$_{\rm rot}$(1$\,$kpc)/\vmax. 

The model population with the DC14 density profile, whose profile is determined by the ratio \mstar/\mhalo,  has galaxies with low values of \vrone/\vmax\, and a large scatter in  the ratio \vrone/\vmax, both in reasonable agreement with observations. The model galaxies with low  \vrone/\vmax\ result from halo expansion, which is included in the DC14 density profile. 
Two effects are responsible for the increased scatter compared with the NFW model: firstly,  the increased fraction of baryonic  mass in the inner regions when profiles are cored, which heightens the effect of the scatter in disc scale lengths. Secondly,  the dependance of the DC14 density profile on   \mstar/\mhalo\  means that halos with the same mass can have different core sizes. 

We note that the model that best fit observations included both halo expansion from non-adiabatic effects of outflows, as well as  adiabatic contraction, both dependent on the ratio of  \mstar/\mhalo. Both baryonic effects  are required to explain the existence of  galaxies  below   {\it and} above  the \vrone\-\vmax\ relation that is predicted by N-body simulations.

We also looked at a specific case of a SIDM model, from which we can draw some general conclusions for models in which core size depends on \mhalo. The  increased fraction of baryonic  mass in the inner regions is also present in models where core size is dependent on halo mass. This means that the scatter  in scale lengths results in such models having significantly more variation in rotation curve shapes than the NFW models. The cores of such models also result in low values of \vrone/\vmax\ in some galaxies, in line with observations.  Scatter in  \vrone/\vmax\ depends on the core size, and can be  as large as  the observations for core sizes of 3\kpc, using the definition of core size of eq.~3.  However, core size may also have a mass dependance in SIDM models  \citep{vogelsberger12}, which is not accounted for in our simple parameterization.   

Further, it is likely that a scatter in core size versus \mhalo\ will exist in SIDM halos, with a dependence on merger  histories. Indeed, \cite{rocha13} find  scatter in core sizes at given halo mass.  Future studies will be able to quantify such scatter, and its mass dependence, which may result in a better match between the SIDM model and observations for smaller core sizes. It is certainly also possible (as recently shown in \citealt{fry15}) that baryonic physics would affect the dark matter density profiles in a SIDM universe, in a mass dependent manner.

 Our study suggests that baryonic physics are able to affect density profiles in dark matter models, to a degree that can account for the observed variation of rotation curve shapes. Such observations may also provide constraints on alternative dark matter models, such as SIDM,  in terms of the scatter and size of cores at a given halo mass. 




\section*{Acknowledgements}
CB acknowledges support through the MINECO grant AYA2012-31101, and the Ramon y Cajal Fellowship.

\bibliographystyle{mn2e}
\bibliography{ms}

\bsp


\label{lastpage}

\end{document}